\documentclass[aps,pre,reprint,superscriptaddress,longbibliography,nobalancelastpage]{revtex4-2}
\usepackage{graphicx}
\usepackage{dcolumn}
\usepackage{amsmath,amssymb}
\usepackage{latexsym}
\usepackage{mathrsfs}
\usepackage[usenames,dvipsnames]{color}
\usepackage{hyperref}
\usepackage{xcolor}
\usepackage{newtxtext}
\usepackage[slantedGreek]{newtxmath}
\usepackage{tikz}
\usepackage{enumitem}

\usetikzlibrary{decorations.pathreplacing,arrows}

\usepackage[normalem]{ulem}
\newcommand{\be}{\begin{equation}} 
\newcommand{\ee}{\end{equation}} 
\newcommand{\bea}{\begin{eqnarray}} 
\newcommand{\eea}{\end{eqnarray}} 

\newcommand{\figref}[2]{[Fig.~\hyperref[#1]{\ref*{#1}(#2)}]}
\newcommand{\figrefi}[2]{[Fig.~\hyperref[#1]{\ref*{#1}(#2)}, inset]}
\newcommand{\textfigref}[2]{Fig.~\hyperref[#1]{\ref*{#1}(#2)}}
\newcommand{\textfigureref}[2]{Figure~\hyperref[#1]{\ref*{#1}(#2)}}
\newcommand{\wholefigref}[1]{(Fig.~\ref{#1})}

\newcommand{\figrefp}[2]{\hyperref[#1]{\ref*{#1}(#2)}}

\definecolor{linkcolor}{HTML}{223096}
\hypersetup{colorlinks,allcolors=linkcolor}

\bibpunct{\textcolor{linkcolor}{[}}{\textcolor{linkcolor}{]}}{\textcolor{linkcolor}{,}}{n}{}{;}
\renewcommand{\eqref}[1]{\hyperref[#1]{(\ref*{#1})}}

\renewcommand{\pi}{\uppi}
\renewcommand{\Delta}{\upDelta}
\DeclareMathAlphabet{\mathcal}{OMS}{cmsy}{m}{n}

\makeatletter
\newcommand{\fmarki}{\ensuremath{\dagger}}
\newcommand{\fmarkii}{\ensuremath{\dagger,\ddagger}}
\newcommand{\fmarkiii}{\ensuremath{\ast}}
    
\def\@fnsymbol#1{{\ifcase#1\or \fmarki\or \fmarkii \or \fmarkiii\else\@ctrerr\fi}}
\makeatother

\begin{document}
\title{Control of lumen morphology by lateral and basal cell surfaces}

\author{Chandraniva \surname{Guha Ray}}
\thanks{These authors contributed equally to this work.}
\affiliation{Max Planck Institute for the Physics of Complex Systems, N\"othnitzer Stra\ss e 38, 01187 Dresden, Germany}
\affiliation{\smash{Max Planck Institute of Molecular Cell Biology and Genetics, Pfotenhauerstra\ss e 108, 01307 Dresden, Germany}}
\affiliation{Center for Systems Biology Dresden, Pfotenhauerstra\ss e 108, 01307 Dresden, Germany}
\author{Markus Mukenhirn}
\altaffiliation[\protect{\phantom{.}\hspace{-5.5mm}\begin{tikzpicture}[scale=0.25]\fill[white](0,0) rectangle (1,1);\end{tikzpicture}\hspace{2.3mm}}Present address:~]{Department of Physiology, Development and Neuroscience, University of Cambridge, Downing Place, Cambridge CB2 3EL, United Kingdom}
\affiliation{\smash{Biotechnology Center, Center for Molecular and Cellular Bioengineering, Technische Universität Dresden, 01069 Dresden, Germany}}
\author{Alf Honigmann}
\email[To whom correspondence should be addressed:\\]{alf.honigmann@tu-dresden.de, haas@pks.mpg.de.}
\affiliation{\smash{Biotechnology Center, Center for Molecular and Cellular Bioengineering, Technische Universität Dresden, 01069 Dresden, Germany}}
\affiliation{Cluster of Excellence Physics of Life, Technische Universität Dresden, 01062 Dresden, Germany}
\author{Pierre A. Haas}
\email[To whom correspondence should be addressed:\\]{alf.honigmann@tu-dresden.de, haas@pks.mpg.de.}
\affiliation{Max Planck Institute for the Physics of Complex Systems, N\"othnitzer Stra\ss e 38, 01187 Dresden, Germany}
\affiliation{\smash{Max Planck Institute of Molecular Cell Biology and Genetics, Pfotenhauerstra\ss e 108, 01307 Dresden, Germany}}
\affiliation{Center for Systems Biology Dresden, Pfotenhauerstra\ss e 108, 01307 Dresden, Germany}
\date{\today}

\begin{abstract}
Across development, the morphology of fluid-filled lumina enclosed by epithelial tissues arises from an interplay of lumen pressure, mechanics of the cell cortex, and cell-cell adhesion. Here, we explore the mechanical basis for the control of this interplay using the shape space of MDCK cysts and the instability of their apical surfaces under tight junction perturbations~[Mukenhirn \emph{et al.}, Dev.~Cell \textbf{59}, 2886 (2024)]. We discover that the cysts respond to these perturbations by significantly modulating their lateral and basal tensions, in addition to the known modulations of pressure and apical belt tension. We develop a mean-field three-dimensional vertex model of these cysts that reproduces the experimental shape instability quantitatively. This reveals that the observed increase of lateral contractility is a cellular response that counters the instability. Our work thus shows how regulation of the mechanics of all cell surfaces conspires to control lumen morphology.
\end{abstract}\maketitle
\renewcommand{\floatpagefraction}{.999}
\section{\uppercase{Introduction}}
The formation of fluid-filled lumina enclosed by epithelial cells~\figref{fig:introduction}{a} is a fundamental process in organ development, enabling fluid transport, nutrient absorption, and secretion in tissues such as the kidney, lung, and intestine. The diversity of the shapes of these lumina is paralleled by the diversity of the mechanical forces, molecular mechanisms, and morphogenetic processes involved in their formation~\cite{Lubarsky2003,andrew10,datta11,leptin14,Navis2015,Navis2016,Camelo2021,torressanchez21,bovyn24}. Physically, the shape of the lumen is controlled by a balance between hydrostatic pressure, tensions in the apical, lateral, and basal cell cortices, cell-cell adhesion at the lateral cell surfaces, and adhesion of the cells to the surrounding extracellular matrix (ECM) at their basal surface [Figs.~\figrefp{fig:introduction}{a} and \figrefp{fig:introduction}{b}]. We review the cell biological origin of these mechanical forces in more detail at the end of this Introduction.

Physical models thus describe lumina as pressurised cavities enclosed by a contractile cell layer~\cite{Gin10,dasgupta18,duclut2019,Vasquez21,duclut21,Turlier21,torressanchez21,bovyn24,mukenhirn24,Lee2024}. Many of these models focus on the interplay of hydraulic and osmotic pressure that drives lumen growth dynamics~\cite{dasgupta18,Turlier21,torressanchez21} and associated hydroelectric effects~\cite{duclut2019,duclut21}. The mechanics of the cells surrounding the lumen often enter these models via a Young--Laplace equation~\cite{degennes} linking the hydraulic pressure to a constant or elastic tension. Other models focus on the mechanics of the apical surface \cite{mukenhirn24}. These models do not resolve the mechanical differences between the apical, lateral, and basal cell surfaces that result of apico-basal polarisation of epithelial cells. The role of the lateral and basal cell surfaces for lumen morphology thus remains less well understood.

A widely used model that can resolve these differences between the cell surfaces is the vertex model \cite{farhadifar07,staple10,fletcher14,fletcher16,alt17}. Motivated both by biological applications and fundamental physical questions, recent work has begun to focus on fully three-dimensional instantiations of this vertex model~\cite{honda04,hannezo14,sanematsu21,sahu21,zhang22,villeneuve24}. In particular, mean-field limits~\cite{bovyn24} could yield insights into the role of the different cell surfaces for lumen morphology.

The role of the different cell surfaces in lumenogenesis also remains poorly understood experimentally, partly because quantifying their mechanics is challenging. Indeed, perturbing cortical tensions or cell-cell adhesion without altering cell mechanics globally is difficult in general~\cite{maitre12}.

\emph{In vitro} models of lumen formation are a means of meeting this challenge: Madin--Darby Canine Kidney (MDCK) cells are widely used to model this process. In three-dimensional culture, MDCK cells polarise and form spherical cysts with a central lumen~\cite{McAteer1986,montesano91,obrien02,Vasquez21,mukenhirn24}, as shown in \textfigref{fig:introduction}{c}. A recent study~\cite{mukenhirn24} introduced tight junction (TJ) perturbations into this system: Loss of TJ scaffolds [ZO-KO, \textfigref{fig:introduction}{d}] led to lumen volume collapse and apical buckling, while cysts lacking TJ strands [CLDN-KO, \textfigref{fig:introduction}{e}] formed rounded lumina despite increased leakiness. Strikingly, the single-cell apical area was observed to be constant in each of these conditions~\figref{fig:introduction}{f}. Less surprisingly perhaps, the single-cell volume is also conserved in each condition~\figref{fig:introduction}{g}. The apical buckling in ZO-KO cysts is associated with strongly reduced lumen pressure, while CLDN-KO cysts retained pressure~\figref{fig:introduction}{h}. TJ loss was found to increase apical belt tension via myosin accumulation~\figref{fig:introduction}{i}. These results pointed to control of apical tension as a key factor in cyst stability~\cite{mukenhirn24}, with TJs maintaining lumen shape by supporting hydrostatic pressure and suppressing apical contractility. 

\begin{figure}[th!]
\centering\includegraphics[width=8.5cm]{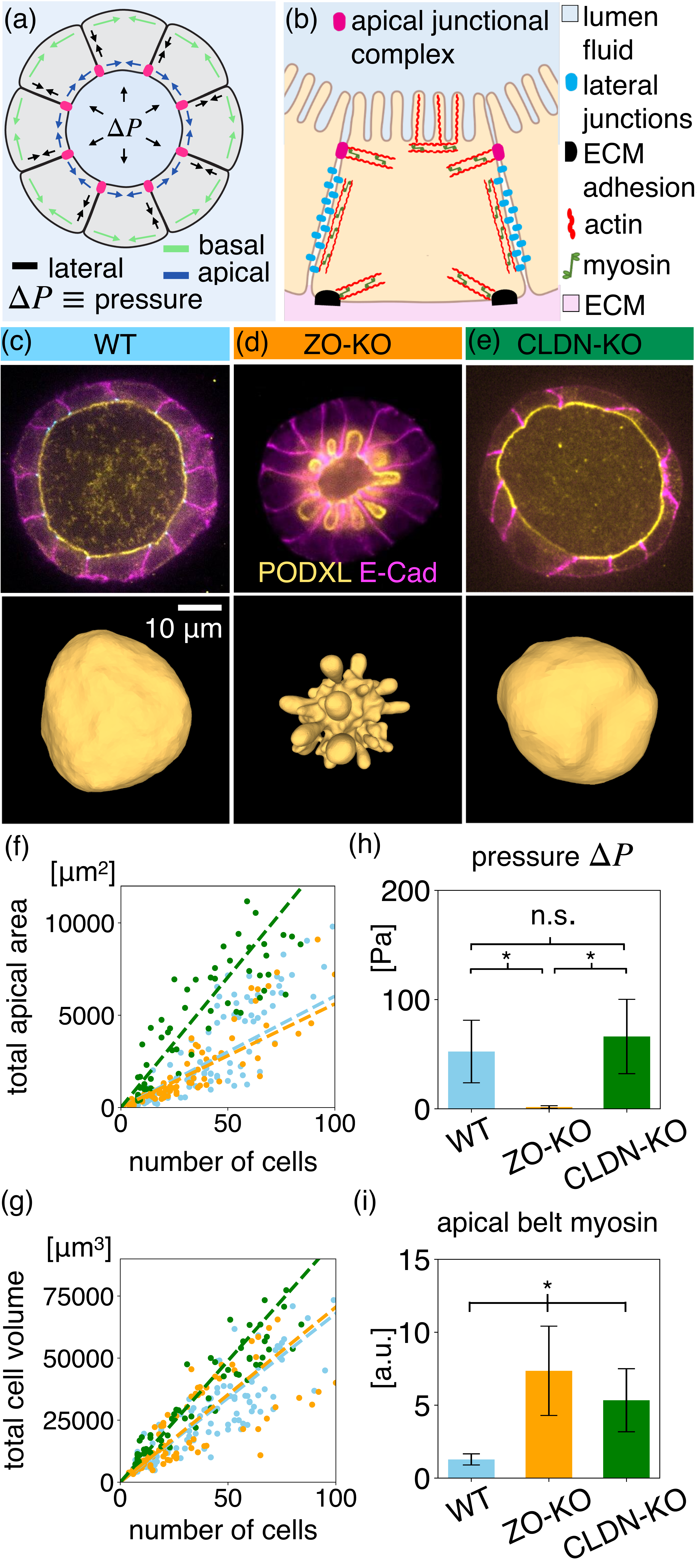}
\end{figure}
\begin{figure}
\vspace{-8pt}
\caption{Mechanics and cell biology of epithelial cyst and behaviour of cysts of Madin--Darby Canine Kidney (MDCK) cells under tight junction (TJ) perturbations~\cite{mukenhirn24}. (a)~Schematic of an epithelial cyst. The lumen is a pressurized, fluid-filled spherical cavity surrounded by a monolayer of polarized epithelial cells. Arrows indicate the hydrostatic pressure difference $\Delta P$ and the cortical tensions acting along the apical, lateral, and basal surfaces, which together determine cyst shape and stability. (b)~Schematic depiction of epithelial cell interfaces and cytoskeletal organization. The apical junctional complex, composed of TJs and adherens junctions, encircles the apex of the cell and anchors the apical actomyosin belt, which generates contractile tension. The lateral membrane forms adhesive contacts between neighbouring cells via cadherin-based adherens junctions. At the basal side, focal adhesions mediate interactions between the cells and the extracellular matrix (ECM) surrounding the cyst. A cortical actin cytoskeleton supports these lateral and basal interfaces and underlies the apical membrane, including actin-rich microvilli. (c)~Midplane confocal sections of a wild-type (WT) MDCK cyst and 3D segmentation of its lumen. Lateral membranes are labelled with E-cadherin (E-Cad, magenta) and apical membranes with podocalyxin (PODXL, yellow). (d)~Corresponding images for a TJ mutant cyst (ZO-KO), showing an instability of the apical cell surfaces that buckle into the cells. (e)~Corresponding images for a different TJ mutant cyst (CLDN-KO); there is no instability of the apical surfaces.  (f)~Quantification of apical surface area from 3D segmentations of lumen and cyst surfaces, plotted against total cell number per cyst in WT ($n=118$ cysts), ZO-KO ($n=87$ cysts), and CLDN-KO ($n=64$ cysts). The dashed straight-line fits indicate that the apical area per cell is constant in each condition. 
(g)~Quantification of the combined volume of all cells in the cyst from 3D cyst segmentations, plotted against total cell number per cyst in WT ($n=118$ cysts), ZO-KO ($n=87$ cysts), and CLDN-KO ($n=64$ cysts). The dashed straight-line fits show that the single-cell volume is constant in each condition. 
(h)~Hydrostatic lumen pressure measurements in MDCK cysts in WT ($n=22$ cysts), ZO-KO ($n=10$ cysts), CLDN-KO ($n=14$ cysts).
(i)~Quantification of myosin enrichment at the apical junctions in WT ($n=125$ junctions), ZO-KO ($n=107$ junctions), and CLDN-KO ($n=92$ junctions). 
$\ast$:~$p<0.01$, n.s.: not significant. Images in panels (c)--(e) reproduced and data in panels (f)--(i) taken from Ref.~\cite{mukenhirn24}.}\label{fig:introduction}
\begin{tikzpicture}[>=stealth]
\draw[<-] (0,0) -- (8.6,0);
\end{tikzpicture}
\end{figure}

Here, we use these MDCK cysts with TJ perturbations to study the role of the lateral and basal cell surfaces for their morphological instability and hence for lumen mechanics more generally. We combine quantitative imaging and a mean-field vertex model of the cyst to reveal that TJ perturbations are accompanied by strong modulations of the mechanics of the lateral and basal cell sides. We show that the increase of the ratio of lateral to basal contractility that we observe stabilises cyst architecture. This suggests that epithelial cells coordinate mechanics across the apical, lateral, and basal surfaces to maintain lumen shape under stress.

\vspace{-8pt}
\subsection*{Cell biology and mechanics of lumina}
We close this Introduction by reviewing the main cell biological contributions to cell and cyst mechanics~[Figs.~\figrefp{fig:introduction}{a} and~\figrefp{fig:introduction}{b}] that we rely on in this paper.
MDCK cells in particular and epithelial cells more generally are polarized along their apico-basal axis~\cite{leptin14,Buckley22}. 
Their mechanical properties are defined by three distinct yet interconnected surface domains~\cite{Buckley22}: the apical, lateral, and basal interfaces~[Figs.~\figrefp{fig:introduction}{a} and~\figrefp{fig:introduction}{b}].

The apical surface faces the lumen and contains an actin-rich cortex including actin-based microvilli that is anchored by the apical junctional complex, composed of tight junctions (TJs) and adherens junctions~\cite{otani20,miyoshi08,Mangeol24}. This complex defines the boundary of the apical domain, regulates paracellular permeability, and can generate contractile tension via the activation of myosin. The lateral surface mediates intercellular adhesion via cadherin-based adherens junctions~\cite{Buckley22,Mangeol24}, while the basal surface anchors the epithelium to the extracellular matrix (ECM) through focal adhesions~\cite{Yu05,Buckley22,OBrien01}. A continuous cortical actin network supports these domains and underlies the apical membrane~\cite{Maraspini20,Roper00}\figref{fig:introduction}{b}.

\begin{figure*}[t!]
\centering 
\includegraphics[width=17.8cm]{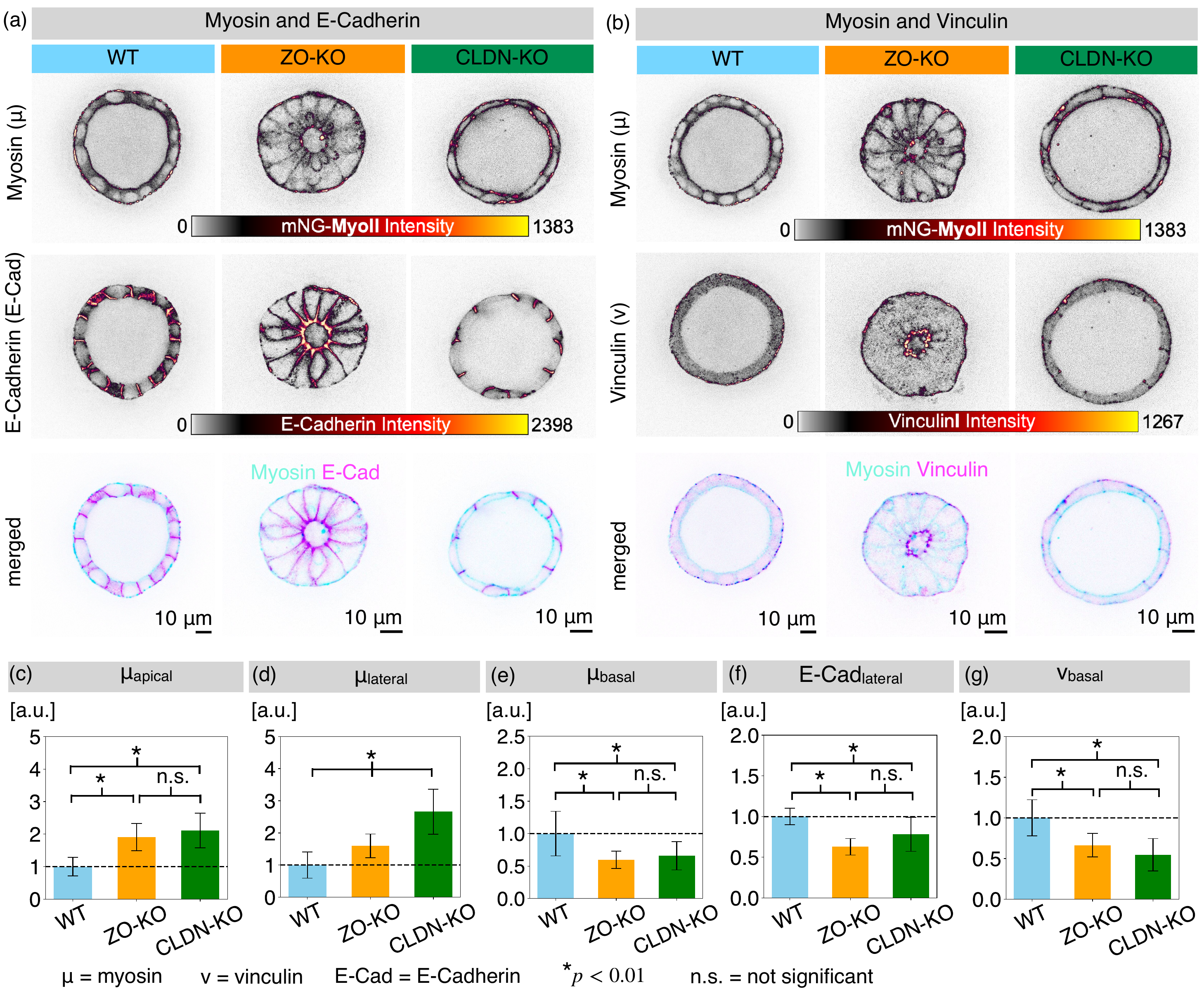}
\caption{Experimental quantification of contractility and adhesion across all cell surfaces wild-type (WT), ZO-KO, and CLDN-KO  MDCK-II cysts. (a) Distribution of myosin IIa--mNeonGreen (top row) and of E-cadherin (middle row) in WT, ZO-KO, CLDN-KO cysts. (b) Distribution of myosin IIa–mNeonGreen (top row) and of vinculin (middle row) in WT, ZO-KO, and CLDN-KO cysts. Fluorescence intensity is shown in arbitrary units [a.u.], visualized using MQ-div-Magma LUT in Fiji~\cite{fiji12}. The bottom rows in panels (a) and (b) show the merged signal. Scale bars: $10\,\text{\textmu m}$.
(c) Quantification of myosin IIa--mNeonGreen intensity (in a.u., chosen so that $\text{WT}=1$) at apical  regions of WT ($n=13$ cysts), ZO-KO ($n=13$ cysts), and CLDN-KO ($n=11$ cysts). (d) Analogous quantification for lateral regions. (e) Analogous quantification for basal regions. (f) Quantification of lateral E-cadherin intensity [a.u.] in WT ($n=13$ cysts), ZO-KO ($n=13$ cysts), and CLDN-KO ($n=11$ cysts). (g) Quantification of basal vinculin intensity (a.u.) in WT ($n=14$ cysts), ZO-KO ($n=15$ cysts), and CLDN-KO ($n=14$ cysts). Data represent fluorescence signal intensities measured from fixed cysts in 3D culture (Appendices~\ref{appA} and~\ref{appB}).}
\label{fig:morphologies}
\end{figure*}

Forces generated at one interface, such as apical belt contractility, can redistribute across the tissue, and basal ECM anchorage can influence overall cell shape \cite{Navis2016,villeneuve24,sahu21}.


TJs are key regulators of this system~\cite{beutel19,mukenhirn24,otani20}: Claudins form the transmembrane strands that limit ion permeability~\cite{otani20}, while \mbox{ZO-1} and {ZO-2} act as cytoplasmic scaffolds linking junctions to the actin cytoskeleton and shaping cortical architecture~\cite{otani20}. 

\section{\uppercase{Experimental results}}
To investigate the role of the lateral and basal cell surfaces for lumen morphology, we analysed how TJs affect all cell surfaces of wild-type (WT), CLDN-KO, and ZO-KO MDCK-II cysts [Figs.~\figrefp{fig:introduction}{a}--\figrefp{fig:introduction}{c} and \ref{fig:morphologies}]. As described above, ZO-KO cysts frequently displayed collapsed or buckled lumina, but CLDN-KO cysts maintained a round morphology similar to WT [Figs.~\figrefp{fig:introduction}{a}--\figrefp{fig:introduction}{c}]. Two mechanical properties of these cysts were quantified in previous work~\cite{mukenhirn24}: Lumen pressure measurements revealed that CLDN-KO cysts retained normal hydrostatic pressure, while ZO-KO cysts were depressurized \figref{fig:introduction}{h}. Quantification of myosin enrichment at the apical junctional belt~\cite{mukenhirn24}, based on 2D monolayers in which apical junctions are easier to resolve, showed the lowest levels in WT, higher levels in CLDN-KO, and the highest levels in ZO-KO cysts \figref{fig:introduction}{i}.

To examine how tight junction perturbations affect the mechanical and structural properties of all cell surfaces, we needed to quantify myosin contractility and adhesion at the apical, lateral, and basal surfaces of 3D cysts \wholefigref{fig:morphologies}. For this, we acquired and analysed new imaging data (Appendices~\ref{appA} and \ref{appB}). In more detail, we quantified the localisation of key cytoskeletal and adhesion markers: myosin IIa (actomyosin contractility), E-cadherin (lateral cell-cell adhesion), and vinculin (basal focal adhesion). Fluorescence intensities were measured across apical, lateral, and basal compartments [Figs.~\figrefp{fig:morphologies}{a} and \figrefp{fig:morphologies}{b}]. Quantification of these signals distinguished between apical and lateral surfaces and the apical belt region (Appendix~\ref{appB}). 

Apical and lateral cortical myosin intensities were higher in CLDN-KO and ZO-KO cysts than in WT [Figs.~\figrefp{fig:morphologies}{c} and \figrefp{fig:morphologies}{d}], indicating increased contractility at these interfaces in the mutants. This is similar to the previously observed apical belt tension in the mutants~\figref{fig:introduction}{i}. By contrast, basal myosin levels~\figref{fig:morphologies}{e} were highest in WT and reduced in both mutant lines. Lateral E-cadherin intensity \figref{fig:morphologies}{f} was strongest in WT, consistent with intact cell-cell adhesion, and decreased in ZO-KO and CLDN-KO cysts. Similarly, basal vinculin intensity \figref{fig:morphologies}{g}, marking integrin-based adhesions to the extracellular matrix, was higher in WT than in ZO-KO or CLDN-KO cysts. Vinculin is a mechanosensitive protein recruited to focal adhesions via talin and adherens junctions via alpha-catenin respectively, where it reflects mechanical force transmission~\cite{le19}. Its strong apical localization in ZO-KO cysts \figref{fig:morphologies}{g}, consistent with previous observations~\cite{otani19}, thus indicates elevated apical belt tension. By contrast, WT cysts show little to no apical vinculin, all of which agrees with the previously reported apical belt tension of WT, ZO-KO and CLDN-KO cysts \figref{fig:introduction}{i}.

These data indicate that tight junction scaffolds not only regulate pressure and tension in the apical surface and the apical belt, but also orchestrate contractility and adhesion across lateral and basal surfaces, which are expected to contribute to the mechanical integrity of epithelial cysts.

\section{\uppercase{Mean-field model of lumen morphology}}
\begin{figure*}[th!]
\centering 
\includegraphics[width=16cm]{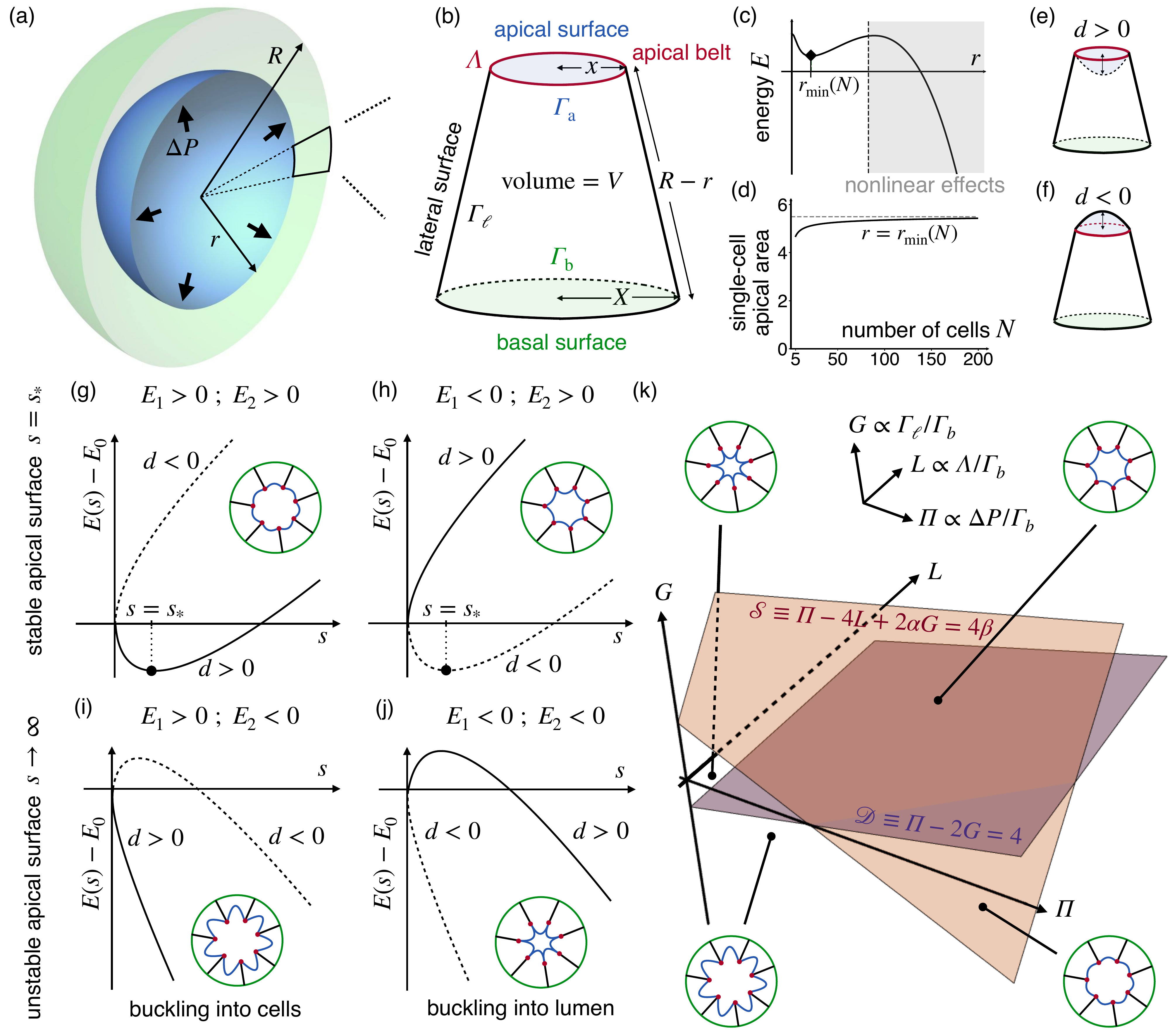}
\caption{Mean-field model of the instability of lumen morphology. (a)~Geometry of a spherical lumen of radius $r$ within a spherical cyst of radius $R$. The pressure difference between lumen and the outside of the cyst is $\Delta P$. (b)~The shape of each single cell is approximated as a truncated cone of apical radius $x$, basal radius $X$, and lateral length $R-r$. The apical, basal, and lateral sides have respective surface tensions $\Gamma_\text{a},\Gamma_\text{b},\Gamma_\ell$, and the apical belt tension is $\Lambda$. The cells are incompressible, and have volume $V$. (c)~Generic plot of the cyst energy $E$ against lumen radius $r$, displaying a local minimum at $r=r_\text{min}(N)$, where $N$ is the number of cells. The unphysical global minimum $r\to\infty$ is regularised by including nonlinear effects (shaded area). Parameter values: $\Gamma_\text{a} = \Gamma_\text{b} = 1, \Gamma_\ell = 20, \Delta P = 0.1, V = 1, N = 200$. (d)~Plot of the apical area $\pi x^2$ of a single cell at the energy minimum $r=r_\text{min}(N)$ against $N$. For low pressure differences, $\Delta P$, the single-cell apical area is approximately constant, and converges as $N\to\infty$ (dotted line). Parameter values as in panel (c), except $\Delta P = 0.01$. (e)~Spherical cap of height $d>0$ buckling into the cell. (f)~Spherical cap of height $d>0$ buckling into the lumen. (g)~Plot of nondimensionalised energy $E(s)$ against the perturbation coordinate $s=r_\ast-r$, in the case $E_1>0$, $E_2<0$, where $r_\ast$ and the coefficients $E_1,E_2$ are defined in the text. The plot shows the branches for $d>0$ (solid line) and $d<0$ (dashed line). There is a single minimum at $s=s_\ast$, with $d>0$: The apical surface is stable and bends slightly into the cells. (h)~Analogous plot for $E_1<0$, $E_2>0$. There is a again a single minimum, now with $d<0$. The apical surface is again stable, but now bends slightly into the lumen. (i)~Analogous plot for $E_1>0$, $E_2<0$. There is no local minimum ($s\to\infty$). The branch with $d>0$ has the lowest energy, so the apical surfaces are unstable and buckle into the cells. (j)~Analogous plot for $E_1<0$, $E_2<0$. Again, there is no local minimum but now the energy is lowest on the branch with $d<0$: The apical surfaces are unstable and buckle into the lumen. (k)~Phase diagram of the buckling instability. The axes are the dimensionless parameters $G\propto\Gamma_\ell/\Gamma_\text{b}$, $L\propto L/\Gamma_\text{b}$, $\Pi\propto\Delta P/\Gamma_\text{b}$ defined in the main text. The planes $\mathscr{S}\equiv\Pi-4L+2\alpha G=4\beta$ and $\mathscr{D}\equiv\Pi-2G=4$, where $\alpha,\beta$ are defined in the text, decide the stability and direction of bending of the apical surfaces and divide parameter space into two stable regions and two unstable regions in which the apical surfaces buckle into the lumen and into the cells, respectively (insets). See text for further explanation.}
\label{fig:model}
\end{figure*}
In order to understand the effect of basal and lateral tension on cyst morphology and the observed lumen instability, we therefore built a mean-field mechanical model of a spherical cyst that resolves the mechanical properties of all cell surfaces. 

The cyst has radius $R$ and consists of a monolayer of $N$ cells surrounding a lumen of radius $r$~\figref{fig:model}{a}. We approximate the shapes of individual cells as truncated circular cones~\figref{fig:model}{b}. Their lateral sides thus have length $R-r$, and the radii of the apical and basal surfaces are $x$ and $X>x$, respectively. The cells are incompressible, of fixed volume $V$. The apical surface areas of the $N$ cells add up to the surface area of the lumen, while their basal surface areas add up to the surface area of the cyst. This means that
\begin{align}
4\pi r^2&=N\pi x^2,&&4\pi R^2=N\pi X^2.\label{eq:area}
\end{align}
Similarly, the volumes of the cells and the volume of the lumen add up to the volume of the cyst, whence
\begin{align}
\frac{4 \pi}{3} R^3 = \frac{4 \pi}{3} r^3 + NV.\label{eq:volume}
\end{align}
We assume that the mechanics of the cells are dominated by cortical tensions, cell-cell adhesion, and cell-matrix adhesion. We adopt a minimal description~\cite{hannezo14,fletcher16,alt17,bovyn24} in which these properties define surface tensions $\Gamma_\text{a},\Gamma_\text{b},\Gamma_\ell$ of the apical, basal, and lateral cell surfaces, and a line tension $\Lambda$ in the apical belt of the cell~\figref{fig:model}{b}. The apical and basal surfaces have areas $\pi x^2$ and $\pi X^2$, respectively; the area of lateral surface is $\pi(RX-rx)$. The apical belt has length $2\pi x$, and so the energy of a single cell is
\begin{align}
e_\text{cell}=\Gamma_\text{a}(\pi x^2)+\Gamma_\text{b}(\pi X^2)+\Gamma_\ell[\pi(RX-rx)]+\Lambda(2\pi x).\label{eq:cellE}
\end{align}
The pressure in the lumen is $P_\text{lumen}$, that in the cell monolayer is $P_\text{cells}$, and the external pressure is $P_0$. An infinitesimal deformation causes infinitesimal volume changes $\delta V_\text{lumen}$ and $\delta V_\text{cyst}$ of the lumen and of the cyst, respectively. Since the cell monolayer is incompressible, $\delta V_\text{cyst}=\delta V_\text{lumen}$. This deformation is therefore resisted by the infinitesimal work
\begin{align}
\delta W&=(P_\text{lumen}-P_\text{cells})\delta V_\text{lumen}+(P_\text{cells}-P_0)\delta V_\text{cyst}\nonumber\\
&=\Delta P\,\delta V_\text{lumen}, 
\end{align}
where $\Delta P=P_\text{lumen}-P_0$. The lumen is pressurised by osmotic imbalances~\cite{torressanchez21}, so $\Delta P>0$. In equilibrium, the cyst minimises its enthalpy, which is therefore
\begin{subequations}
\begin{align}
E=Ne_\text{cell}+\Delta P\,V_\text{lumen}.\label{eq:enthalpy}
\end{align}
Explicitly, on substituting Eqs.~\eqref{eq:area} and \eqref{eq:volume} into Eq.~\eqref{eq:cellE},
\begin{align}
\dfrac{E}{4\pi}&=-\dfrac{\Delta P}{3}r^3+\Gamma_\text{a}r^2+\Gamma_\text{b}\left(r^3+\dfrac{3NV}{4\pi}\right)^{2/3}\nonumber\\
&\qquad+\dfrac{\Gamma_\ell}{2}\sqrt{N} \left[\left(r^3 + \dfrac{3NV}{4\pi}\right)^{2/3}-r^2\right]+\Lambda\sqrt{N}r.\label{eq:E0}
\end{align}
\end{subequations}
This energy has a global minimum as $r\to\infty$ \figref{fig:model}{c}. This minimum is unphysical and would be regularised by nonlinear effects such as the mechanics of the cell nucleus and the confinement energy of the cytoplasm~\cite{hannezo14} that are not included in our model. Generically, the energy has a local minimum at $r=r_\ast(N)$ that determines the lumen radius.
\subsection{Conservation of single-cell apical area}
This model can provide a mechanical explanation for the experimental observation that the apical area of a single cell is conserved in spherical cysts as the number $N$ of cells increases. Indeed, the single-cell apical area is $A=4\pi r_\ast(N)^2/N$ from the first of Eqs.~\eqref{eq:area}, and \textfigref{fig:model}{d} illustrates that that there are parameter values for which $A$ changes very little as $N$ varies within its experimentally relevant range.

\begin{widetext}\enlargethispage{1cm}
To explain this observation, we note that $A=\text{const.}$ requires $r_\ast(N)\sim\sqrt{N}$. We change variables by introducing $\eta=r/\sqrt{N}$. To prove that $A=\text{const.}$ is now to prove that $\eta$ is independent of $N$ at the energy minimum. Equation~\eqref{eq:E0} becomes
\begin{subequations}
\begin{align}
\dfrac{E}{4\pi N}&=-\dfrac{\Delta P}{3}\sqrt{N}\eta^3+\Gamma_\text{a}\eta^2+\Gamma_\text{b}\eta^2\left(1+\dfrac{3V}{4\pi\sqrt{N}\eta^3}\right)^{2/3}+\dfrac{\Gamma_\ell}{2}\sqrt{N}\eta^2\left[\left(1+\dfrac{3V}{4\pi\sqrt{N}\eta^3}\right)^{2/3}-1\right]+\Lambda\eta,
\end{align}
\end{subequations}
\end{widetext}
whence\setcounter{equation}{5}
\begin{subequations}
\begin{align}\setcounter{equation}{1}
\dfrac{E}{4\pi N}&\sim-\dfrac{\Delta P}{3}\sqrt{N}\eta^3+\left[(\Gamma_\text{a}+\Gamma_\text{b})\eta^2+\dfrac{\Gamma_\ell V}{4\pi\eta}+\Lambda\eta\right],
\end{align}    
\end{subequations}
for $N\gg 1$. If $\Delta P$ is sufficiently small, the energy minimum is determined by the terms in square brackets only. Now the single-cell volume is independent of $N$~\figref{fig:introduction}{g}. Hence so is $\eta$, assuming that the (relative) tension magnitudes do not change with $N$. Thus single-cell apical area is indeed conserved in this limit.

In the converse limit, in which $\Delta P$ is very large, the energy has a single minimum with $r\to\infty$. This would imply that $R-r\to 0$, which is resisted by the incompressibility of the cell nuclei. Hence, in this limit, $R-r=\varrho$, where $\varrho$ is the size of the cell nuclei. Thus
\begin{align}
\varrho=\left(r^3+\dfrac{3NV}{4\pi}\right)^{1/3}-r\sim \dfrac{NV}{4\pi r^2}\;\Longrightarrow\; r\sim\left(\dfrac{NV}{4\pi \varrho}\right)^{1/2},
\end{align}
for $N\gg 1$. Since $V$ is independent of $N$~\figref{fig:introduction}{g}, this exhibits the scaling $\smash{r\sim\sqrt{N}}$, so the single-cell apical area is conserved in this limit, too.

All of this shows that the experimentally observed conservation of single-cell apical area in spherical cysts \emph{can} emerge purely mechanically, but this argument cannot of course exclude active regulation of single-cell apical area, nor can it explain conservation of single-cell apical area in the non-spherical ZO-KO lumina.

\subsection{Instability of lumen morphology}
To describe the instability of the apical surfaces in these ZO-KO cysts, we extend our model by replacing the flat apical surfaces of the truncated-cone shapes in \textfigref{fig:model}{b} with spherical caps of height $d$~[Figs.~\figrefp{fig:model}{d} and \figrefp{fig:model}{e}]: if $d>0$, the spherical cap points into the cells, while it points into the lumen if $d<0$. The spherical cap has volume $\pi|d|(3x^2+d^2)/6$. Using the first of Eqs.~\eqref{eq:area}, Eq.~\eqref{eq:volume} is therefore replaced with
\begin{subequations}
\begin{align}
\dfrac{4\pi}{3}R^3=\left[\dfrac{4\pi}{3}r^3+\dfrac{N\pi}{6}d\left(\dfrac{12r^2}{N}+d^2\right)\right]+NV,
\end{align}
in which the term in square brackets is the volume of the lumen. This rearranges to
\begin{align}
R=\left(r^3+\dfrac{3dr^2}{2}+\dfrac{Nd^3}{8}+\dfrac{3NV}{4\pi}\right)^{1/3}.\label{eq:R}
\end{align}
\end{subequations}
The single-cell apical area is now $A=\pi(x^2+d^2)$. In line with experimental observations, we will now assume that $A$ is conserved, so that
\begin{align}
d=\pm\left(a-\dfrac{4r^2}{N}\right)^{1/2},\label{eq:d}
\end{align}
in which $a=A/\pi$ is therefore constant, and where we have used, again, the first of Eqs.~\eqref{eq:area}.
\subsubsection{Energy of the cyst}
With these geometric results and the assumption of single-cell apical area conservation, Eq.~\eqref{eq:cellE} is replaced with
\begin{align}
e_\text{cell}&=\Gamma_\text{a}A+\Gamma_\text{b}(\pi X^2)+\Gamma_\ell[\pi(RX-rx)]+\Lambda(2\pi x)\nonumber\\
&=\text{const.}+\dfrac{4\pi \Gamma_\text{b}}{N}R^2+\dfrac{2\pi\Gamma_\ell}{\sqrt{N}}\bigl(R^2-r^2\bigr)+\dfrac{4\pi\Lambda}{\sqrt{N}}x,
\end{align}
using Eqs.~\eqref{eq:area}. On discarding the constant single-cell apical area and substituting into the expression for the enthalpy of the cyst in Eq.~\eqref{eq:enthalpy}, and on nondimensionalising lengths with $\sqrt{a}$ and energies with $4\pi\Gamma_\text{b}a$, we find
\begin{align}
E&=-p\left(\dfrac{r^3}{3}+\dfrac{dr^2}{2}+\dfrac{Nd^3}{24}\right)+R^2+\dfrac{\gamma}{2}\sqrt{N}\bigl(R^2-r^2\bigr)+\lambda r\sqrt{N},
\end{align}
where we have defined
\begin{align}
&p=\dfrac{\sqrt{a}\,\Delta P}{\Gamma_\text{b}},&&\gamma=\dfrac{\Gamma_\ell}{\Gamma_\text{b}},&&\lambda=\dfrac{\Lambda}{\Gamma_\text{b}\sqrt{a}},\label{eq:params2}
\end{align}
and where, from Eqs.~\eqref{eq:R} and \eqref{eq:d},
\begin{subequations}
\begin{align}
d&=\pm\left(1-\dfrac{4r^2}{N}\right)^{1/2},\label{eq:d2}\\
R&=\left[r^3\pm\left(1-\dfrac{4r^2}{N}\right)^{1/2}\left(r^2+\dfrac{N}{8}\right)+Nv\right]^{1/3},
\end{align}
\end{subequations}
in which we have defined $v=3V/4\pi a^{3/2}$. Again, this model does not include cell compressibility (which the experimental data suggest to be negligible) and nonlinear cell mechanical effects, associated for example with the cell nuclei. It does not include the bending energy of the apical surface either. The latter may not be negligible for highly buckled shapes, but can be neglected at the onset of the instability which we seek to describe here.
\subsubsection{Stability analysis}
If $d=0$, then, from Eq.~\eqref{eq:d2}, $r=r_\ast\equiv\sqrt{N}/2$. For $d\neq 0$, Eq.~\eqref{eq:d2} implies that $r<r_\ast$. We analyse the stability of the apical surfaces by expanding the energy close to $d=0$, i.e., for $s\equiv r_\ast-r\ll 1$. This leads to
\begin{align}
E=E_0\mp E_1s^{1/2}+E_2s+O\bigl(s^{3/2}\bigr),\label{eq:eexp}
\end{align}
in which
\begin{subequations}\label{eq:E1E2}
\begin{align}
E_1&=\dfrac{N^{5/12}}{4\bigl(\sqrt{N}+8v\bigr)^{1/3}}\left\{p\bigl[N\bigl(\sqrt{N}+8v\bigr)\bigr]^{1/3}-2\sqrt{N}\gamma-4\right\},\\
E_2&=\dfrac{Np}{4}-\sqrt{N}\lambda-\dfrac{N^{1/6}}{\bigl(\sqrt{N}+8v\bigr)^{1/3}}\left(\sqrt{N}+\dfrac{1}{\sqrt{N}+8v}\right)\nonumber\\
&\quad+\dfrac{N\gamma}{2}\left\{1-\dfrac{1}{\bigl[N\bigl(\sqrt{N}+8v\bigr)\bigr]^{1/3}}\left(\sqrt{N}+\dfrac{1}{\sqrt{N}+8v}\right)\right\}.
\end{align}
\end{subequations}
An expression for $E_0$ can be obtained similarly, but is of no further interest, as it only sets a reference state of the energy. There are now four possible behaviours \mbox{[Figs.~\figrefp{fig:model}{g}--\figrefp{fig:model}{j}]}, depending on the signs of $E_1$ and $E_2$: If $E_2>0$, the energy has a minimum, either on the branch with $d>0$ [$E_1>0$, \textfigref{fig:model}{g}] or on the branch with $d<0$ [$E_1<0$, \textfigref{fig:model}{h}]. In both cases, the apical surface is stable; it bends slightly into the cells in the first case \figrefi{fig:model}{g} and into the lumen in the second case \figrefi{fig:model}{h}. If $E_2<0$, there is no energy minimum on either branch, and the apical surfaces are unstable. If, additionally, $E_1>0$, then the branch with $d>0$ has the lowest energy \figref{fig:model}{i}, and the apical surfaces buckle into the cells \figrefi{fig:model}{i}, while, if $E_1<0$, then the lowest-energy branch is the one with $d<0$ \figref{fig:model}{j}, and so the apical surfaces buckle into the lumen \figrefi{fig:model}{j}.

The minimum for $E_2>0$ discussed above is attained at $s=s_\ast\equiv(E_1/2E_2)^2$. Extending the expansion in Eq.~\eqref{eq:eexp} to $E=E_0\mp E_1s^{1/2}+E_2s\mp E_3s^{3/2}+O\bigl(s^2\bigr)$, asymptoticity demands $\smash{E_2s_\ast\gg E_3s_\ast^{\smash{3/2}}}$, i.e., $\smash{E_2^2\gg E_1E_3}$. This condition holds and hence the asymptotic expansion is valid in the limit $N\gg1$: using \textsc{Mathematica}, we find
\begin{align}
\dfrac{E_2^2}{E_1E_3}\sim -\dfrac{6p^2}{(p-2\gamma)(19\gamma-14p)}N+O\bigl(N^{1/2}\bigr)\gg 1.
\end{align}

\subsubsection{Phase diagram of lumen morphology}
The phase diagram of the instability thus depends on the signs of $E_1$ and $E_2$. Now
\begin{subequations}\label{eq:planes}
\begin{align}
E_1\gtrless 0\quad&\Longleftrightarrow\quad \Pi-2G-4\gtrless0,\label{eq:plane1}\\
E_2\gtrless 0\quad&\Longleftrightarrow\quad \Pi-4L+2\alpha G-4\beta\gtrless 0,\label{eq:plane2}
\end{align}
\end{subequations}
where, from Eqs.~\eqref{eq:E1E2},
\begin{align}
\Pi&=p\bigl[N\bigl(\sqrt{N}+8v\bigr)\bigr]^{1/3},&G&=\sqrt{N}\gamma,&L&=\dfrac{\bigl(\sqrt{N}+8v\bigr)^{1/3}}{N^{1/6}}\lambda,\label{eq:params}
\end{align}
and
\begin{subequations}\label{eq:alphabeta}
\begin{align}
\alpha&=\dfrac{\bigl(\sqrt{N}\!+\!8v\bigr)^{1/3}}{N^{1/6}}\left\{1-\dfrac{1}{\bigl[N\bigl(\sqrt{N}\!+\!8v\bigr)\bigr]^{1/3}}\left(\sqrt{N}+\dfrac{1}{\sqrt{N}\!+\!8v}\right)\right\},\label{eq:alpha}\\
\beta&=1+\dfrac{1}{\sqrt{N}\bigl(\sqrt{N}+8v\bigr)}\label{eq:beta},
\end{align}
\end{subequations}
In particular, $\alpha\sim 8v/3\sqrt{N}$ and $\beta\sim 1$ for $N\gg 1$. The planes $\Pi-2G=4$ and $\Pi-4L+2\alpha G=4\beta$ divide $(\Pi,L,G)$ parameter space into four regions~\figref{fig:model}{k}. This calculation shows that these regions corresponding to the four possibilities discussed above [\textfigref{fig:model}{k}, insets]. In particular, the plane $\mathscr{D}\equiv\Pi-2G=4$ determines the direction of bending of the apical surfaces (into the lumen or into the cells), while the plane $\mathscr{S}\equiv\Pi-4L+2\alpha G=4\beta$ determines the stability of the apical surfaces.

\subsubsection{Mechanisms of lumen morphology}
Equation~\eqref{eq:plane2} and \textfigref{fig:model}{k} show that higher values of $\Pi-4L+2\alpha G$ favour stability. Thus, high $\Pi$ or $G$ are stabilising and high~$L$ is destabilising. Indeed, at high $L$ (i.e., high belt tension), the apical belt contracts, so the apical surface must buckle to maintain constant area. This contraction reduces the radius of the lumen, which is resisted by high $\Pi$ (i.e., high pressure), which is therefore stabilising. Similarly, high~$G$ (i.e., high lateral tension) shrinks the lateral cell sides, so the lumen radius must increase by cell volume conservation, which stabilises the apical surfaces.

Similarly, Eq.~\eqref{eq:plane1} and \textfigref{fig:model}{k} show that higher values of $\Pi-2G$ drive bending or buckling of the apical surface into the cells. Thus, unsurprisingly, the apical surfaces bend or buckle into the cells at high $\Pi$ (i.e., at high pressure). Since the lateral cell sides shrink at high $G$ (i.e., high lateral tension), volume conservation favours bending or buckling of the apical surfaces into the lumen. More surprisingly perhaps, the direction of bending or buckling is independent of $L$, i.e., of apical belt tension. This independence emerges from the full calculation, but it is tempting to explain it by noting that the direction of bending or buckling as the apical belt contracts does not affect apical area conservation at leading order.

\section{Discussion}
To test our theoretical model and to understand the mechanical role of the different cell surfaces for lumen morphology, we combine our model with the experimental measurements.

\subsection{MDCK cysts in model parameter space}
We begin by quantifying the experiments in the parameter space of the stability diagram. To leading, linear order, any tension $T$ in our model can be expressed as  ${T = k \upmu - k' \text{a}}$, where $\upmu$ and $\text{a}$ are the fluorescence intensities of contractile proteins (myosin) and adhesive proteins (E-cadherin or vinculin) respectively, and $k$ and $k'$ are proportionality constants that are, in general, different for different cell surfaces (Appendix~\ref{appB}). 

Since we have nondimensionalised our model parameters with the basal tension, these nondimensional parameters can be expressed in terms of the experimental measurements in Figs.~\figrefp{fig:morphologies}{c}--\figrefp{fig:morphologies}{g} rescaled with basal myosin intensities.

\begin{figure}[t!]
\centering\includegraphics[width=8.5cm]{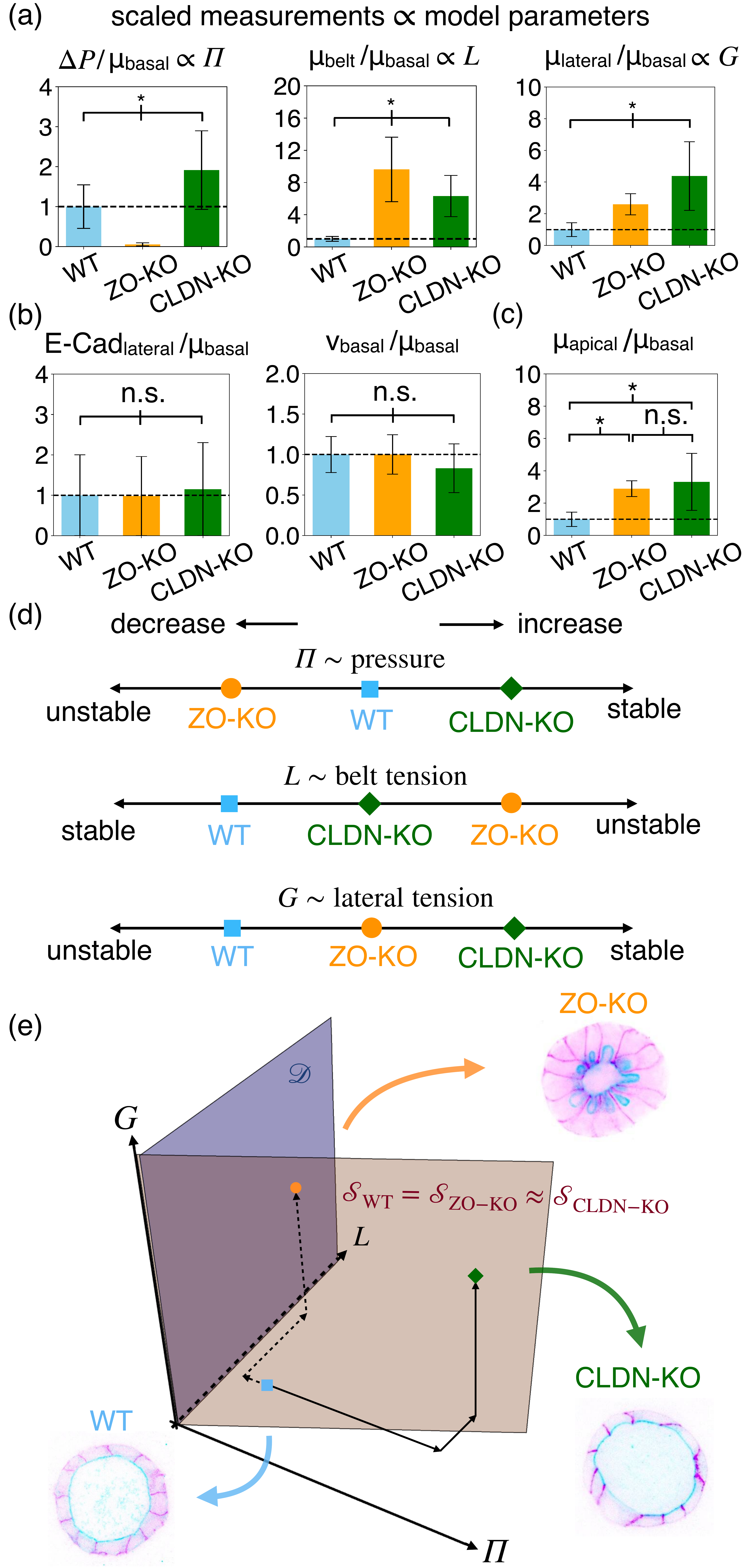}
\end{figure}

\begin{figure}
\vspace{-8pt}
\caption{Experimental quantification of the instability of lumen morphology. (a)~Plots of the ratios $\Delta P/\upmu_\text{basal}$, $\upmu_\text{belt}/\upmu_\text{basal}$, $\upmu_\text{lateral}/\upmu_\text{basal}$ that determine the parameters $\Pi,L,G$ of the phase diagram of the instability~\figref{fig:model}{k}, for WT, ZO-KO, and CLDN-KO cysts. All three ratios change significantly ($\ast$: $p<0.01$). (b)~Similar plots of the ratios $\text{Ecad}_\text{lateral}/\upmu_\text{basal}$, $\text{v}_\text{basal}/\upmu_\text{basal}$, showing that they do not change significantly (n.s.) between WT, ZO-KO, and CLDN-KO conditions. (c)~Plot of the ratio $\upmu_\text{apical}/\upmu_\text{basal}$, which does change significantly between WT and ZO-KO or CLDN-KO, but does not enter the model because of the experimental observation that single-cell apical area is conserved~\figref{fig:introduction}{f}. Plots in panels (a)--(c) are based on the data in Figs.~\figrefp{fig:introduction}{h}, \figrefp{fig:introduction}{i}, and \figrefp{fig:morphologies}{c}--\figrefp{fig:morphologies}{g}, and values are scaled with the WT values (so that $\text{WT}=1$). (d) Effect of the parameters $\Pi,L,G$ on stability, illustrating the stabilising or destabilising effects of the changes of each of these parameters, between WT, ZO-KO, and CLDN-KO conditions, observed in panel (a). (e)~Phase diagram, analogous to \textfigref{fig:model}{k}, showing a set of values of $\Pi,L,G$ consistent with the observed cyst morphologies and the experimental measurements in panel (a). See text for further explanation. Arrows indicate the changes of $\Pi,L,G$ for $\text{WT}\to\text{ZO-KO}$ and $\text{WT}\to\text{CLDN-KO}$ transitions [based on panels (a) and (d)]; insets show WT, ZO-KO, and CLDN-KO cysts for illustration.}\label{figure4}
\begin{tikzpicture}[>=stealth]
\draw[<-] (0,0) -- (8.6,0);
\end{tikzpicture}
\end{figure}

The pressure, apical belt myosin intensity, and lateral myosin intensity still change significantly between WT and the perturbations if they are rescaled with the basal myosin intensity in this way~\figref{figure4}{a}. By contrast, the changes of the lateral E-Cadherin and basal vinculin intensities are compensated by changes of the basal myosin intensity, so the corresponding rescaled intensities do not change significantly~\figref{figure4}{b}. 

We also note that the ratio of apical to basal myosin intensities, which does not enter our model founded on the experimentally observed conservation of apical areas~\figref{fig:introduction}{f}, does change significantly between WT and the two perturbations. These changes may be related to active regulation of the single-cell apical area.

As explained in Appendix~\ref{appB}, these rescaled measurements allow us to estimate ratios or differences of model parameters. In particular, we obtain the changes, between WT, ZO-KO, and CLDN-KO cysts, of the parameters $\Pi,L,G$, the effect of which on lumen morphology we have discussed above. As shown in \textfigref{figure4}{d},
\begin{itemize}[leftmargin=*,itemsep=0pt]
\item the pressure parameter $\Pi$ increases, i.e., stabilises, from ZO-KO to WT to CLDN-KO;
\item the belt tension parameter $L$ increases, i.e., destabilises, from WT to CLDN-KO to ZO-KO;
\item the lateral tension parameter $G$ increases, i.e., stabilises, from WT to ZO-KO to CLDN-KO.
\end{itemize}
The destabilising response of the apical belt to both TJ perturbations was already highlighted in Ref.~\cite{mukenhirn24}. Strikingly however, our analysis also reveals that the increase of the lateral tension parameter $G$ counteracts the instability in both TJ perturbations. This may indicate that the lateral cell sides are regulated to stabilise apical surfaces and lumen morphology.

The destabilising decrease of the pressure parameter $\Pi$ in the ZO-KO is hardly surprising. Unexpectedly however, $\Pi$ increases in the CLDN-KO in spite of increased leakiness. Importantly, the increased pressure difference in the CLDN-KO compared to WT~\figref{fig:introduction}{h} is not statistically significant. The significant increase in $\Pi$~\figref{figure4}{a} rather results from the statistically significant decrease in basal myosin levels~\figref{fig:morphologies}{e}. This hints at regulation of the basal cell surfaces to stabilise the apical surfaces and lumen morphology.

Since the proportionality constants between tensions and intensities remain undetermined, we can only estimate ratios of the parameters $\Pi,L,G$, but not their absolute values (Appendix~\ref{appB}). Still, we can make our comparison between model and experiment more quantitative: there exist values of the parameters $\Pi,L,G$ that that are consistent with these measured ratios and also with the observed stability behaviour. An instance of this is given by the set of parameter values plotted in the phase diagram in \textfigref{figure4}{e}.

\begin{figure}[t!]
\centering\includegraphics[width=8.5cm]{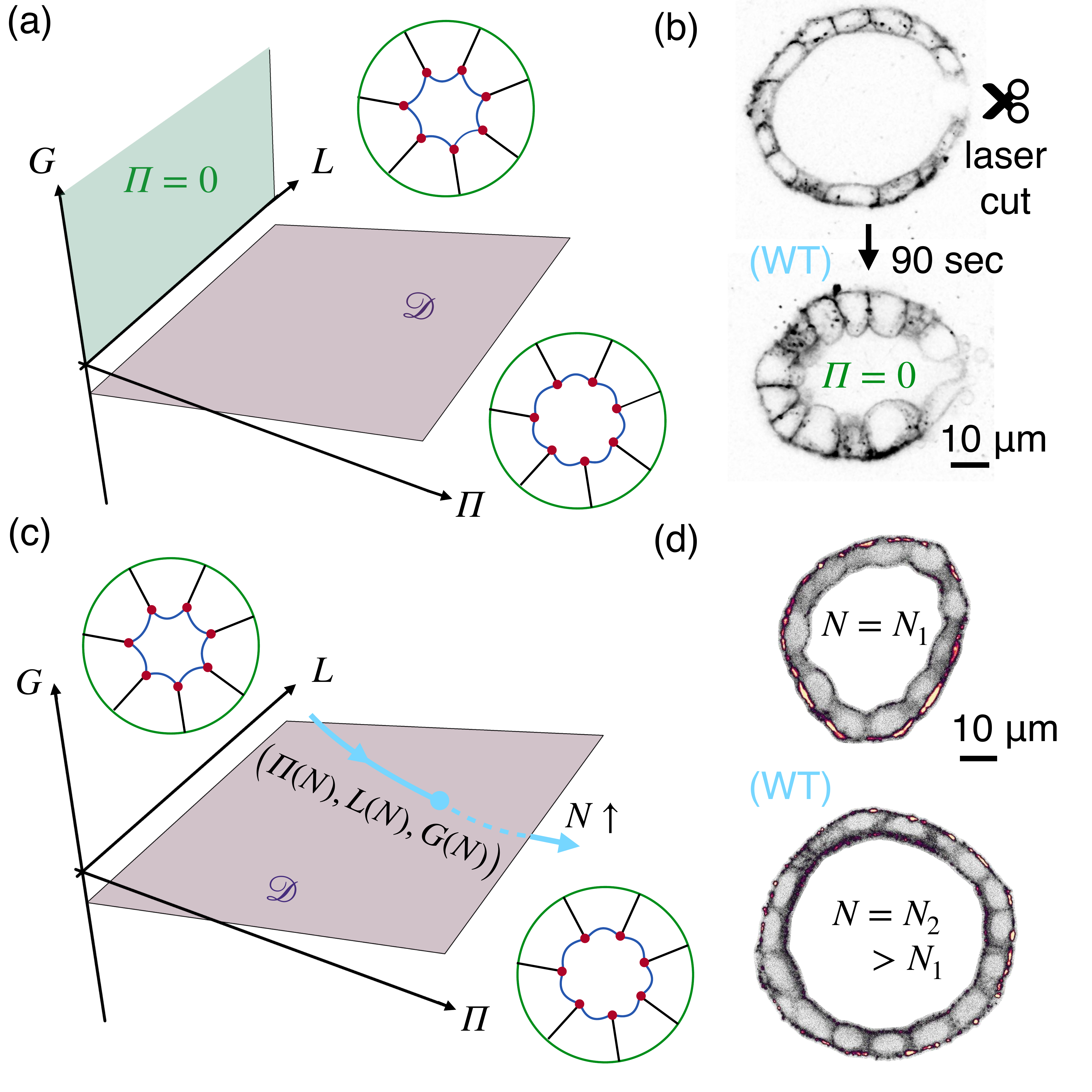}
\caption{Additional model predictions. (a)~At zero pressure ($\Pi=0$) and for contractile lateral sides $(G>0)$, the model predicts that the apical cell surfaces bend or buckle into the lumen (shape insets on either side of the plane $\mathscr{D}$ determining the direction of bending). (b)~Laser cuts opening cysts set $\Delta P=0$ and lead to bending of apical cell surfaces into the lumen in WT. Images taken from Ref.~\cite{mukenhirn24}. (c)~As the number $N$ of cells increases, $(\Pi,L,G)$ trace out a path in parameter space that can cross $\mathscr{D}$, so the model allows the apical surfaces of ``small'' cysts to bend into the lumen, while those of ``large'' cysts bend into the cells. (d)~Instances of a WT cyst with $N=N_1$ cells and apical cell surfaces bent into the lumen and of a WT cyst with $N=N_2>N_1$ cells and apical cell surfaces bent into the cells. Scale bars: $10\,\text{\textmu m}$.}\label{fig5}
\end{figure}

\subsection{Additional model predictions}
Our mean-field model makes two more testable predictions about the direction of bending or buckling of apical surfaces.

First, cysts with $\Pi=0$ (corresponding to zero pressure difference) and $G>0$ (corresponding to contractile rather than extensile lateral sides) lie above the plane $\mathscr{D}\equiv\Pi-2G=4$ in model parameter space. Hence their apical surfaces will bend or buckle into the lumen \figref{fig5}{a}. This prediction can be tested by laser cuts opening the cyst~\cite{mukenhirn24} that equalise hydrostatic pressures instantaneously, $\Delta P=0$. In the WT~\figref{fig5}{b} and consistently with the predictions of the model, this causes the apical surfaces that were initially bent into the cells to bend or buckle into the lumen. This inversion is observed, too, in the CLDN-KO, but not in the ZO-KO, as illustrated in Video S2 of Ref.~\cite{mukenhirn24}. It is likely that nonlinear effects prevent inversion of the buckled apical surfaces in the ZO-KO.

Second, for fixed values of the dimensionless pressure $p$, apical belt tension $\lambda$, and lateral tension $\gamma$ defined in Eq.~\eqref{eq:params2}, $\bigl(\Pi(N),L(N),G(N)\bigr)$ trace out a curve in $(\Pi,L,G)$ parameter space as the number $N$ of cells increases, given explicitly by Eqs.~\eqref{eq:params}. It can start out above the plane~$\mathscr{D}$ at small $N$, then cross below $\mathscr{D}$ at larger $N$~\figref{fig5}{c}. This predicts that the apical surfaces of ``small'' cysts with few cells bend into the lumen, while those of ``large'' cysts with more cells bend into the cells. This agrees with observations \figref{fig5}{d}. Whether changes of $p,\lambda,\gamma$, assumed constant to arrive at this prediction, also contribute to this observed change of the bending direction of the apical surfaces, remains an open question.

\subsection{Conclusions}
In this paper, we have studied the cell mechanical basis for lumen morphology using MDCK cysts, relying on tight junction mutations to perturb cyst mechanics and morphology. Our mean-field mechanical model can reproduce the buckling instability of ZO-KO cysts quantitatively. This model predicts the onset of the instability, but not the buckled shapes, which are determined by unknown nonlinear mechanisms. Our combined experimental and theoretical results reveal that lumen shape is also set by lateral and basal tensions, and not only by apical belt tension and pressure. Moreover, our results indicated the tensions in the lateral and basal cell surfaces are regulated to stabilise lumen shape. Thus the mechanics of all cell surfaces combine to control lumen morphology. Elucidating the cell biological basis for these observations remains, however, an open problem.

Quantifying the fluorescence intensities of contractile and adhesive proteins allowed us to estimate ratios of cortical tensions across WT and the perturbations (Appendix~\ref{appB}), but did not yield absolute values of these tensions because of the unknown prefactors relating intensities to tensions. Indeed, we relied on the laser ablation experiments of Ref.~\cite{Guyomar24} to decide that contractility dominates adhesion in the lateral and basal cell sides. These prefactors could in principle be determined by simultaneous measurement of intensities and quantitative tension measurements using ablation or atomic force microscopy experiments~\cite{Guyomar24,fischer16}. However, these prefactors could be different in different cell surfaces~\cite{Guyomar24}, and the experimental challenge is compounded by the fact that existing experimental approaches often rely on protein overexpression. It is for this reason that absolute measurements that resolve contractile and adhesive contributions to the tensions in the different cell surfaces remain a formidable challenge for future work.

More physically, we have shown that the observed conservation of single-cell apical area during lumen growth~\figref{fig:introduction}{f} can emerge passively in different parameter limits of our model. Whether the system relies on this passive mechanism or uses active mechanisms such as balancing apical endo- and exocytosis~\cite{Kamalesh21,Simoes22} remains, however, an unanswered question. Mechanical nonlinearities, such as the bending energy of the apical surface, could also contribute to a passive mechanism for apical area conservation, especially in the strongly buckled apical surfaces of ZO-KO cysts.
\begin{acknowledgments}
We thank Matt Bovyn and Mike Staddon for discussions or comments on the manuscript and Cécilie Martin-Lemaitre for cell-culture work. We also thank the light-microscopy facility of MPI-CBG and the light-microscopy facility of CMCB Dresden for help and support with imaging. C.G.R. and P.A.H. were supported by the Max Planck Society; M.M. and A.H. were supported by the Volkswagen Foundation (project number A133289).
\end{acknowledgments}

\section*{Data availability}
Example raw imaging data and the data underlying the quantifications in Figs.~\ref{fig:morphologies} and \ref{figure4} are openly available at Ref.~\footnote{Data are available at \href{https://doi.org/10.5281/zenodo.17055494}{\texttt{zenodo:17055495}}}. Full raw imaging data are available from the corresponding authors on reasonable request.

\appendix
\section{\uppercase{Experimental Methods}}\label{appA}
\subsection{MDCK cell culture and maintenance}
Madin--Darby Canine Kidney-II (MDCK) cell lines with tight junction perturbations used for this study were as follows: the ZO1/2 knockout MDCK-II (ZO-KO) cell line was generated by Beutel \emph{et al.}~\cite{beutel19} using CRISPR/Cas9, and the Claudin quin-KO (CLDN-KO) MDCK-II cell line was obtained from the Furuse lab~\cite{otani19}. Myosin II was endogenously labeled in these cell lines as described in Ref.~\cite{mukenhirn24}. Briefly, N-terminal endogenous labelling of Myosin II-A with mNeon in WT, ZO-KO, and CLDN-KO cells was achieved by targeting the MYH9 exon using CRISPR/Cas.
All cell lines were cultured in Dulbecco's Modified Eagle Medium (DMEM) supplemented with 5\% fetal bovine serum (FBS) and 1\% penicillin-streptomycin. Cells were maintained at 37°C in a humidified atmosphere with 5\% CO\textsubscript{2}. Media were refreshed every 2--3 days, and cells were passaged at 80--90\% confluence.

\subsection{3D culture of MDCK cysts}
MDCK cysts were cultured following established protocols. Briefly, MDCK monolayers were dissociated and embedded in 50\% Matrigel (Corning, USA). Cells were maintained for 5~days before immunofluorescence staining.

\subsection{Immunofluorescence staining}
Cells were seeded onto glass coverslips and cultured until they reached the desired confluence. They were then fixed with 4\% paraformaldehyde for 15 minutes at room temperature, permeabilised with 0.1\% Triton X-100 for 15 minutes, and blocked with 5\% bovine serum albumin (BSA) for 1~hour. Primary antibodies against Vinculin (Sigma-Aldrich, \#V4505) and E-Cadherin (Cell Signaling, \#3195) were applied overnight at 4°C. After washing, fluorescently labelled secondary antibodies were applied for 1 hour at room temperature. Images were acquired using a spinning disk confocal microscope.

\section{\uppercase{Data analysis}}\label{appB}
\subsection{Analysis of experimental images}
Apical junctional myosin in 2D monolayers was quantified using previously published data~\cite{mukenhirn24}. In this dataset, endogenous myosin IIa tagged with mNeonGreen was imaged in MDCK-II WT and mutant monolayers. To quantify apical belt myosin, the average fluorescence intensity at apical cell–cell junctions was measured and compared between genotypes. Regions of interest (ROIs) were manually drawn along the apical junctions in z-projected confocal images, and mean intensities were extracted using Fiji \cite{fiji12}.

3D quantification of myosin intensity in cysts was performed on confocal z-stacks of MDCK cysts stained for endogenous myosin IIa, E-Cadherin (lateral marker), and Vinculin (basal marker). To quantify average myosin intensity at specific membrane domains, hand-drawn ROIs were placed at the apical surface, lateral membrane, and basal side in individual optical sections. Care was taken to exclude the apical junctional complex from the lateral ROI by restricting each ROI to its corresponding subcellular domain. For example, both apical and lateral ROIs were drawn to avoid the apical belt region, ensuring separation of apical and lateral myosin signals. Average fluorescence intensities within each region were measured, and background signal was subtracted using a nearby extracellular region. In what follows, ``intensity'' refers to these averaged data corrected for background for signal. All image analysis was performed in Fiji \cite{fiji12}.
\subsection{Estimation of model parameters}
These measurements allow us to estimate or constrain the three model parameters $\Pi$, $G$, and $L$, and the two parameters $\alpha$ and $\beta$ that define the slopes of the planes in the phase diagram in \textfigref{figure4}{e}, for the WT and the two perturbations, ZO-KO and CLDN-KO.
\subsubsection{Estimation of tensions}
We begin by estimating the surface and line tensions from the measured intensities of myosin, E-cadherin and vinculin. To linear order, the tensions are directly proportional to the intensity of contractile proteins (mainly myosin $\upmu$), and negatively proportional to the intensity $\text{a}$ of adhesive proteins [E-cadherin ($\text{e}$) and vinculin ($\text{v}$)]. In general, a tension $T$ can therefore be written as
\begin{align}
T = k \upmu - k' \text{a},
\end{align}
in which the proportionality constants $k,k'>0$ relate arbitrary intensity units to tensions, and will depend both on the mechanical properties of the protein as well as the cortex. Hence these constants are in general different at  different cell surfaces~\cite{Guyomar24}. Here, we will find that this does not affect the ratios of physical parameters that we compare across mutant perturbations. 

The surface tensions of the apical, basal, and lateral surfaces, and the line tension in the apical belt are thus given by
\begin{subequations}\label{eq:tensions}
\begin{align}
\Gamma_{\rm a} &= k_{\rm a} \upmu_{\rm a}, &\Gamma_{\rm b} &= k_{\rm b} \upmu_{\rm b} - k'_{\rm b} {\rm v}_{\rm b},\\
\Lambda &= k_{\rm belt} \upmu_{\rm belt},&\Gamma_{\ell} &= k_{\ell} \upmu_{\ell} - k'_{\ell} {\rm e}_{\ell},
\end{align}
\end{subequations}
where $\upmu_{\rm a}$, $\upmu_{\rm b}$, and $\upmu_{\ell}$ are the respective myosin intensities of the apical, basal, and lateral surfaces, $\upmu_{\rm belt}$ is the myosin intensity of the apical belt, ${\rm v}_{\rm b}$ is the intensity of vinculin on the basal surface, and ${\rm e}_\ell$ is the intensity of E-Cadherin on the lateral surface.

Since our final model parameters $\Pi$, $L$, and, $G$ are nondimensionalised with respect to the basal surface tension, we shall do the same with all of our measurements. Rearranging Eqs.~\eqref{eq:tensions}, 
\begin{align}\label{eq:lbtensions}
\Gamma_{\rm b} &= k_{\rm b}\upmu_{\rm b} \left(1- \dfrac{k'_{\rm b}{\rm v}_{\rm b}}{k_{\rm b}\upmu_{\rm b}} \right),&\Gamma_{\ell} & = k_{\rm b}\upmu_{\rm b} \left(\dfrac{k_{\ell}\upmu_{\ell}}{k_{\rm b}\upmu_{\rm b}}- \dfrac{k'_{\ell}{\rm e}_{\ell}}{k_{\rm b}\upmu_{\rm b}} \right).
\end{align}
Now we leverage the experimental observation that the normalised intensities of E-cadherin and vinculin~\figref{figure4}{b} do not change significantly across the WT and the perturbations. Thus
\begin{align}
&c_1 = 1- \dfrac{k'_{\rm b}{\rm v}_{ \rm b}}{k_{\rm b}\upmu_{\rm b}},&&c_2=\dfrac{k'_{\ell} \mathrm{e}_{\ell}}{k_{\rm b} \upmu_b}
\end{align}
are constants, across the WT and the perturbations. With this, Eqs.~\eqref{eq:lbtensions} become
\begin{align}
\Gamma_{\rm b} &= k_{\rm b}\upmu_{\rm b} c_1, &\Gamma_{\ell} & = k_{\rm b}\upmu_{\rm b} \left(\dfrac{k_{\ell} \upmu_{\ell}}{k_{\rm b} \upmu_{\rm b}}- c_2 \right).\label{eq:Gammasbl}
\end{align}
The recoil of the lateral and basal sides on laser ablation observed in Ref.~\cite{Guyomar24} shows that $\Gamma_\text{b},\Gamma_\ell>0$ in the WT. In particular, $c_1>0$.
\subsubsection{Estimation of parameters of the phase diagram}
We can now estimate (relations between) the model parameters $\Pi,G,L$, which are defined in Eqs.~\eqref{eq:params2} and~\eqref{eq:params}. From Eqs.~\eqref{eq:Gammasbl}, we find
\begin{align}
\Pi&= \frac{\sqrt{a} \Delta P}{k_{\rm b} \upmu_{\rm b} c_1} \bigl[N\bigl(\sqrt{N}+8v\bigr)\bigr]^{1/3},
\end{align}
in which $\Delta P$ is given in \textfigref{fig:introduction}{h}, $N=73\pm33$ is the number of cells in the cysts analysed, and $a=A/\pi$, $v=3V/(4\pi a^{3/2})$, defined above, depend additionally on the constant single-cell apical area $A$ and volume $V$, given in Table~\ref{table:singlecell}, based on the data in Figs.~\figrefp{fig:introduction}{f} and~\figrefp{fig:introduction}{g}. This allows estimation of $\Pi$ in WT and CLDN-KO and ZO-KO cysts up to the unknown constant $k_{\rm b} c_1$. Since $c_1>0$, we have $\Pi^\text{WT},\Pi^\text{ZO},\Pi^\text{CLDN}>0$. The independent ratios $\Pi^\text{WT}/\Pi^\text{ZO},\Pi^\text{ZO}/\Pi^\text{CLDN}$ do not depend on this constant, however, and their values are reported in Table~\ref{table:ratios}. All of this implies that $0<\Pi^\text{ZO}<\Pi^\text{WT}<\Pi^\text{CLDN}$ \figref{figure4}{d}.

\begin{table}[t!]
\caption{Measured values of the single-cell apical area $A$ and single-cell volume $V$ for WT, ZO-KO, and CLDN-KO cysts. Values are obtained from the slopes of the dashed lines in Figs.~\figrefp{fig:introduction}{f} and \figrefp{fig:introduction}{g}. Data taken from Ref.~\cite{mukenhirn24}.}
\begin{ruledtabular}
\begin{tabular}{cccc}
parameter&WT&ZO-KO&CLDN-KO\\
\hline
$A$ $\left[\text{\textmu m}^2\right]\vphantom{A^{\frac{A}{A}}}$&$60\pm33$&$56\pm22$&$141\pm61$\\
$V$ $\left[\text{\textmu m}^3\right]\vphantom{A^{\frac{A}{A}}}$&$679\pm 247$&$705 \pm 273$&$978\pm 261$
\end{tabular}
\end{ruledtabular}
\label{table:singlecell}
\end{table}

\begin{table}[b!]
\caption{Estimates of ratios of parameters determining cyst stability and buckling direction for the phase diagram in \textfigref{figure4}{e}.}
\begin{ruledtabular}
\begin{tabular}{cc}
\hphantom{AAAAA}parameter ratio\hphantom{AAAAA}&\hphantom{AAAAA}estimated value\hphantom{AAAAA}\\
\hline\vspace{6pt}
$\dfrac{\Pi^{\rm WT}\vphantom{A^{\frac{A}{A}}}}{\Pi^{\rm ZO}}$&  $19 $\\
\vspace{6pt}
$\dfrac{\Pi^{\rm ZO}}{\Pi^{\rm CLDN}}$&$0.019$\\
\vspace{6pt}
$\dfrac{G^{\rm WT} - G^{\rm ZO}}{G^{\rm ZO}-G^{\rm CLDN}}$&$0.89$\\
\vspace{6pt}
$\dfrac{L^{\rm WT}}{L^{\rm ZO}}$&$0.097$\\
\vspace{3pt}
$\dfrac{L^{\rm ZO}}{L^{\rm CLDN}}$&$2.93$
\end{tabular}
\end{ruledtabular}
\label{table:ratios}
\end{table}

Using Eqs.~\eqref{eq:params2}, \eqref{eq:params}, and \eqref{eq:Gammasbl} again, the second model parameter, $G$, can be written as
\begin{align}
G = \frac{\sqrt{N} }{c_1} \left(\dfrac{k_{\ell} \upmu_{\ell}}{k_{\rm b} \upmu_{\rm b}}- c_2 \right).
\end{align}
Here, computing the ratios across the WT or the perturbations would still leave us with an unknown offset $c_2$. This offset results from the contribution of adhesion, due to E-cadherin, to the lateral tension. In Table~\ref{table:ratios}, we still report a quantity independent of both $c_1$ and $c_2$, viz.,
\begin{align}
    \dfrac{G^{\rm WT} - G^{\rm ZO}}{G^{\rm ZO}-G^{\rm CLDN}}.
\end{align}
The fact that $\Gamma_\text{b},\Gamma_\ell>0$ in the WT shows that $G^\text{WT}>0$. We can further compute $c_1\bigl(G^\text{WT}-G^\text{ZO}\bigr),\,c_1\bigl(G^\text{ZO}-G^\text{CLDN}\bigr)<0$. As noted above, $c_1>0$, whence $0<G^\text{WT}<G^\text{ZO}<G^\text{CLDN}$ \figref{figure4}{d}. 

The third model parameter is $L$, which, from Eqs.~\eqref{eq:params2}, \eqref{eq:params}, and \eqref{eq:Gammasbl}, can be written as
\begin{align}
L = \dfrac{\bigl(\sqrt{N}+8v\bigr)^{1/3}k_{\rm belt} \upmu_{\rm belt}}{N^{1/6}\sqrt{a} k_{\rm b} \upmu_{\rm b} c_1 }. 
\end{align}
Since $c_1>0$, $L^\text{WT},L^\text{ZO},L^\text{CLDN}>0$. We compute the ratios $L^\text{WT}/L^\text{ZO},L^\text{ZO}/L^\text{CLDN}$ between perturbations once again, which are independent of the unknown constant $k_{\rm b} c_1/k_\text{belt}$ and which we report in Table~\ref{table:ratios}. These results imply that $0<L^\text{WT}<L^\text{CLDN}<L^\text{ZO}$ \figref{figure4}{d}.

\begin{table}[th!]
\caption{Estimates of the parameters $\alpha,\beta$ determining the slopes of the phase diagram in \textfigref{figure4}{e}.}
\begin{ruledtabular}
\begin{tabular}{cccc}
parameter&WT&ZO-KO&CLDN-KO\\
\hline
$\alpha$&$0.41\pm0.28$&$0.45\pm0.23$&$0.19\pm0.12$\\
$\beta$&$1.0049\pm0.0028$&$1.0044\pm0.0021$&$1.0079\pm0.0023$
\end{tabular}
\end{ruledtabular}
\label{table:alphabeta}
\end{table}

The stability diagram also depends on the slope parameters $\alpha$ and $\beta$. We compute their values, reported in Table~\ref{table:alphabeta}, using the definitions in Eqs.~\eqref{eq:alphabeta}. The corresponding stability planes $\mathscr{S}_\text{WT},\mathscr{S}_\text{ZO-KO},\mathscr{S}_\text{CLDN-KO}$, plotted in \textfigref{figure4}{e}, are indistinguishable.

\subsubsection{Identification of points in the phase diagram}
The stability diagram is determined by the values of the three model parameters for each of WT, ZO-KO, and CLDN-KO, i.e., nine model parameters in total. The experimental quantification has yielded five equality constraints involving these, given in Table \ref{table:ratios}. In addition, the observed  stability and buckling direction of the apical surfaces and Eqs.~\eqref{eq:planes} yield six inequality constraints between these parameters. They are
\begin{subequations}
\begin{align}
&\Pi^{\rm WT} - 2 G^{\rm WT} - 4>0,  \\
&\Pi^{\rm ZO} - 2 G^{\rm ZO} - 4>0, \\
&\Pi^{\rm CLDN} - 2 G^{\rm CLDN} - 4>0, \\
&\Pi^{\rm WT} - 4 L^{\rm WT} + 2 \alpha^{\rm WT} G^{\rm WT}- 4\beta^{\rm WT}>0,\\
&\Pi^{\rm ZO} - 4 L^{\rm ZO} + 2 \alpha^{\rm ZO} G^{\rm ZO}- 4\beta^{\rm ZO}<0,\\
&\Pi^{\rm CLDN} - 4 L^{\rm CLDN} + 2 \alpha^{\rm CLDN} G^{\rm CLDN}- 4\beta^{\rm CLDN}>0,
\end{align}
\end{subequations}
in which the values of the slope parameters are given in Table~\ref{table:alphabeta}. Additionally, we have the inequality constraints
\begin{subequations}
\begin{align}
0&<\Pi^\text{ZO}<\Pi^\text{WT}<\Pi^\text{CLDN},\\
0&<G^\text{WT}<G^\text{ZO}<G^\text{CLDN},\\
0&<L^\text{WT}<L^\text{CLDN}<L^\text{ZO}.
\end{align}    
\end{subequations}
obtained above. We find parameter values that satisfy all these equality and inequality constraints using the \texttt{FindInstance} function of \textsc{Mathematica} and plot them in the phase diagram in \textfigref{figure4}{e}. 

\bibliography{main}
\end{document}